\documentclass[aps,pre,groupedaddress,showkeys,twocolumn,longbibliography]{revtex4-2}

\usepackage{amsmath}
\usepackage{graphicx}
\usepackage{float,physics}
\usepackage{siunitx}
\usepackage{xcolor}
\usepackage{epstopdf}
\definecolor{darkgreen}{rgb}{0,0.6,0.0}

\usepackage{latexsym}
\usepackage{amsmath}
\usepackage{amsfonts}
\usepackage{amssymb}
\usepackage{mathrsfs}
\usepackage{verbatim}
\usepackage{anysize}
\usepackage{soul}

\usepackage[utf8]{inputenc}
\usepackage{algorithmic}
\usepackage{algorithm}
\usepackage{enumerate}
\usepackage{amssymb, amsmath, amsthm}
\usepackage{graphicx}
\usepackage{lmodern,url}
\usepackage{graphicx}
\usepackage{wrapfig}
\usepackage{hyperref}

\floatname{algorithm}{Algoritmo}

\newcommand{\bo}{\raise-1mm\hbox{\Large$\Box$}}

\usepackage{multirow}

\begin{document}
	\title{Sequential epidemic spread between agglomerates \\of self-propelled agents in one dimension\\}
	\normalsize
	
	\author{Pablo de Castro}
	\email[]{pablo.castro@ictp-saifr.org}
	\affiliation{ICTP-South American Institute for Fundamental Research - Instituto de F\'isica Te\'orica da UNESP, Rua Dr.~Bento Teobaldo Ferraz 271, 01140-070 S\~ao Paulo, Brazil.}
	
		\author{Felipe Urbina}
	\affiliation{Vicerrector\'ia de Investigaci\'on, Universidad Mayor, Santiago, Chile}
	
		\author{Ariel Norambuena}
	\affiliation{Vicerrector\'ia de Investigaci\'on, Universidad Mayor, Santiago, Chile}
	
	\author{Francisca Guzm\'an-Lastra}
	\affiliation{Department~of~Physics, Faculty of Sciences, Universidad de Chile, Santiago, Chile}
	
%


\date{\today}

\begin{abstract}
	
	Motile organisms can form stable agglomerates such as cities or colonies. In the outbreak of a highly contagious disease, the control of large-scale epidemic spread depends on factors like the number and size of agglomerates, travel rate between them, and disease recovery rate. While the emergence of agglomerates permits early interventions, it also explains longer real epidemics. In this work, we study the spread of susceptible-infected-recovered epidemics in one-dimensional spatially-structured systems. By working in one dimension, we mimic microorganisms in narrow channels and establish a necessary foundation for future investigation in higher dimensions. We employ a model of self-propelled particles which spontaneously form multiple clusters. As the rate of stochastic reorientation decreases, clusters become larger and less numerous. Besides examining the time evolution averaged over many epidemics, we show how the final number of ever-infected individuals depends non-trivially on single-individual parameters. In particular, the number of ever-infected individuals first increases with the reorientation rate since particles escape sooner from clusters and spread the disease. For higher reorientation rate, travel between clusters becomes too diffusive and the clusters too small, decreasing the number of ever-infected individuals.


\end{abstract}
\maketitle

\section{Introduction}

A central problem in epidemiology is how space and individual motion affect epidemic spread \cite{conner2004movement,berestycki2023epidemic}. More broadly, this question concerns many dynamical contexts where individual states are transmitted by contact, whether among animals \cite{cosner2009effects, vasconcelos2021standard, coutinho2021model,jorge2021assessing}, insects \cite{brown2020metacommunity,richardson2017short}, microorganisms \cite{li2020formation}, or other sorts of agents \cite{paoluzzi2020information}. Scenarios where individuals are not homogeneously distributed in space are particularly challenging \cite{martcheva2015spatial}. For instance, motile organisms may form stable agglomerates such as cities or colonies. In fact, the occurrence of diseases typically happens in spatial clusters. As a result, if one can anticipate further spreading, containment becomes feasible \cite{riley2007large}. The existence of a spatial hierarchy of measles transmission in England and Wales was demonstrated through the study of population patches arranged in a line \cite{grenfell2001travelling}. A more sophisticated version of this technique illustrates that excess human mortality related to pneumonia and influenza in the United States is consistent with human travel patterns \cite{viboud2006synchrony}. In the presence of clustering, epidemic spread should be controlled by the typical size of the agglomerates, their typical distance to one another, the rate at which individuals travel between them, and the disease recovery rate. Exactly how these quantities control epidemic spread is an understudied subject. 

In movement epidemiology, increasing attention has been paid to models of self-propelled particles where epidemic spread emerges from individual motion rules \cite{rodriguez2019particle, ghosh2022surface,rodriguez2022epidemic,kim2020coupling,kim2021kinetic,namilae2017self,levis2020flocking}. One example is the active Brownian particle (ABP) model, where the particle moves with a self-propulsion velocity of constant magnitude and whose direction fluctuates stochastically and continuously \cite{rojas2021fast,PhysRevE.107.014608,romanczuk2012active,teixeira2021single}. This model has been largely used to model animal movement, particularly bacteria \cite{romanczuk2012active}. ABPs can spontaneously agglomerate into clusters even if subject to purely repulsive short-range interactions \cite{redner}. Such motility-induced clustering arises as particles block each other due to low reorientation rate, i.e., persistent motion. A stationary cluster size distribution (CSD) is reached once the rates of particle absorption and emission from the clusters become equal \cite{peruani2010cluster}. The ratio between self-propulsion speed and reorientation rate can be interpreted as a proxy for the individual tendency of agglomerating. With motility-induced clustering, no social interactions are required; the clustering is determined and maintained purely via self-propulsion.

Epidemic spread among infectious ABPs was recently studied assuming a susceptible-infected-recovered (SIR) scenario in two spatial dimensions (2D) without clustering, focussing on the role of individual-based (or ``microscopic'') parameters \cite{norambuena2020understanding}.
More recently, the SIR contagion dynamics of self-propelled particles in 2D with repulsive interactions and polar alignment was investigated numerically \cite{zhao2022contagion}. Emerging spatial structures, such as bands and clusters, were shown to strongly affect the final fraction of ever-infected (recovered) individuals.

A simpler model qualitatively similar to ABPs \cite{cates2013active} is the run-and-tumble particle (RTP) \cite{andrea2020,de2021diversity}. In this model, the agent simply moves in a straight line (`run') up to a random time instant where a new random run direction is chosen (`tumble'). RTPs are also frequently used to model bacterial motion \cite{liang2018evaluation,PhysRevE.107.034605,figueroa20203d}. For narrowly confined systems such as bacteria which live in channel-like soil pores, a number of theoretical difficulties are eliminated by assuming a one-dimensional (1D) scenario \cite{de2021active}. In particular, one can derive a parameter-free expression for the exponentially-decaying CSD of RTPs expressed in terms of motion parameters and total density \cite{soto2014run}. 
	 
\begin{figure}[!h]
	\centering
	\includegraphics[width=\columnwidth]{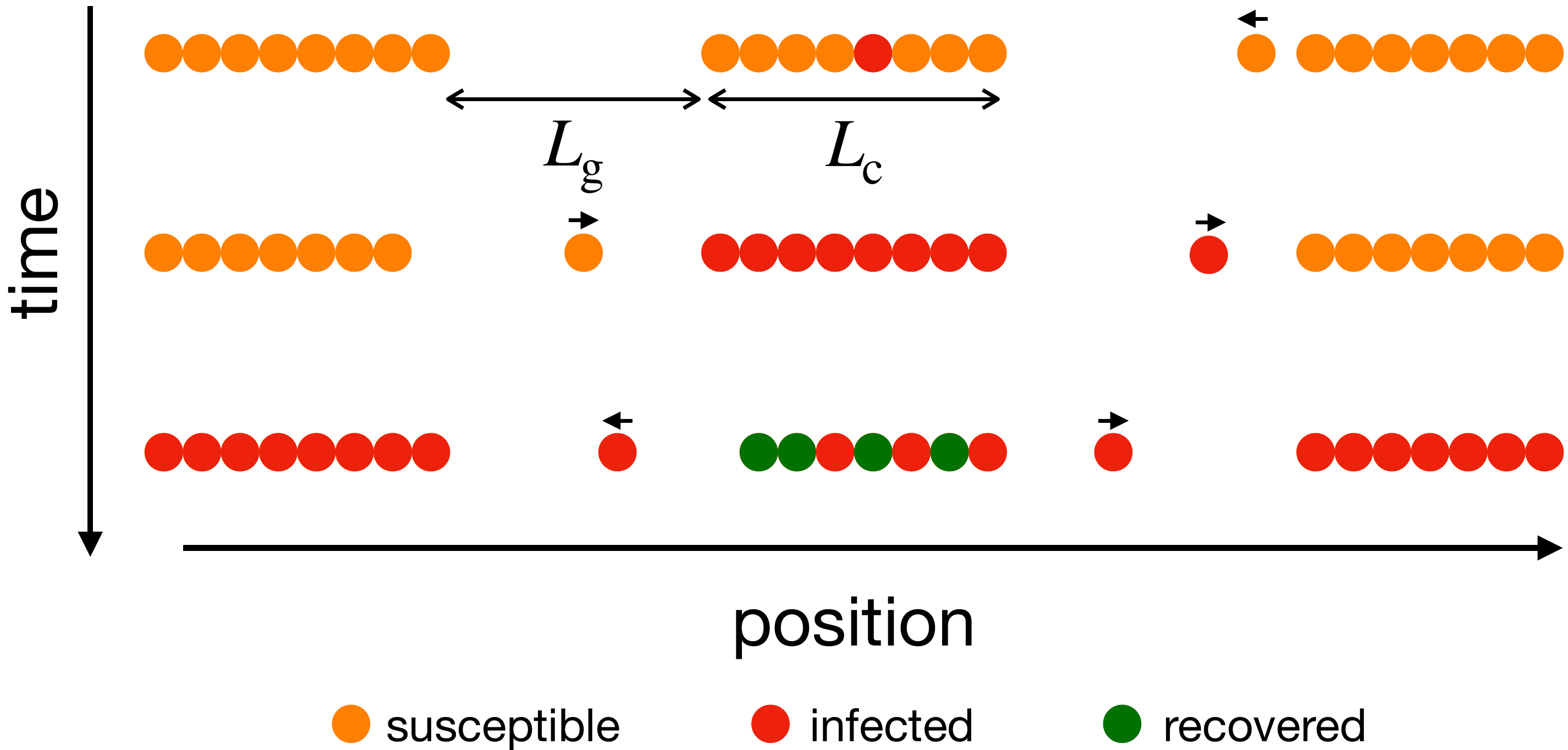}%
	\caption{Schematic zoom for the space-time evolution of the large-scale epidemic spread between clusters. $L_\text{c}$ and $L_\text{g}$ correspond to the average cluster and gas region sizes, respectively. The color scheme is as follows: white corresponds to empty regions, orange is for susceptible individuals, red for infected, and green for recovered. The three configurations shown are not consecutive in time but do follow the time arrow indicated on the left. The tiny arrows over the travelling particles represent their self-propulsion directions.}
	\label{fig:test}
\end{figure}

In the present work, we investigate the large-scale spread of SIR epidemics across a spatial array of multiple agglomerates in one dimension (see Fig.~\ref{fig:test} for a schematic close-up). 
For this purpose, we use a computationally cheap model of RTPs on a 1D discrete lattice \cite{soto2014run}. Motility-induced clustering arises spontaneously as time progresses. The epidemic starts only after clustering becomes stationary. Particle alignment is absent, isolating the role of motility-induced clustering. Tackling the 1D version of this problem is important in three fronts. First, it emulates real-world narrowly confined systems. Secondly, it assesses the \textit{sequential} contribution to contact-transmission spreading between agglomerates. Thirdly, it paves the way towards the intricate 2D version, generating intuition on which features may be relevant. In this direction, recently, 2D simulations were used to study clustering effects on the contagion dynamics of RTPs \cite{forgacs2022using}. Differently than there, here we focus on the role of individual motion parameters and total density as well as on the long-time effects of cluster-cluster travels across multiple clusters. Finally, in our 1D RTP-SIR model, we can explain simulation data analytically, a crucial distinction from previous works.
	

This paper is organized as follows. Section \ref{sec:clustering} defines our RTP model and reviews analytical CSD results, without contagion. Section \ref{sec:SIRdynamics} introduces the SIR dynamics and presents simulation results. A phenomenological theory explaining the data is derived. Section \ref{sec:micro} derives, validates, and discusses a microscopic parameter-free theory for the final fraction of ever-infected individuals. A second and more accurate version of the latter theory is then described, where the input clustering observables do not come from another theory but rather from the simulations directly. Finally, Section \ref{sec:micro} brings conclusions.

\section{Run-and-tumble clustering}
\label{sec:clustering}
We start by reviewing the clustering model (without contagion) used in the rest of the paper \cite{soto2014run}. A discrete periodic lattice in 1D is considered, with ${N=2000}$ sites, $M$ particles, and maximal occupancy of one particle per site. Each particle has a director (self-propulsion direction), towards left or right. The dimensionless total concentration is denoted by $\phi \equiv M/N$. Within each time step, $M$ particles are selected randomly and sequentially (particle repetition is allowed but rare). With probability $\nu$, the director is redrawn. If the director points to an empty neighboring site, the particle moves into it. The initial state corresponds to random positions and directors. The free-particle self-propulsion speed fluctuates around $v=1$. 
	\begin{figure}[!h]
	\centering
	\includegraphics[width=\columnwidth]{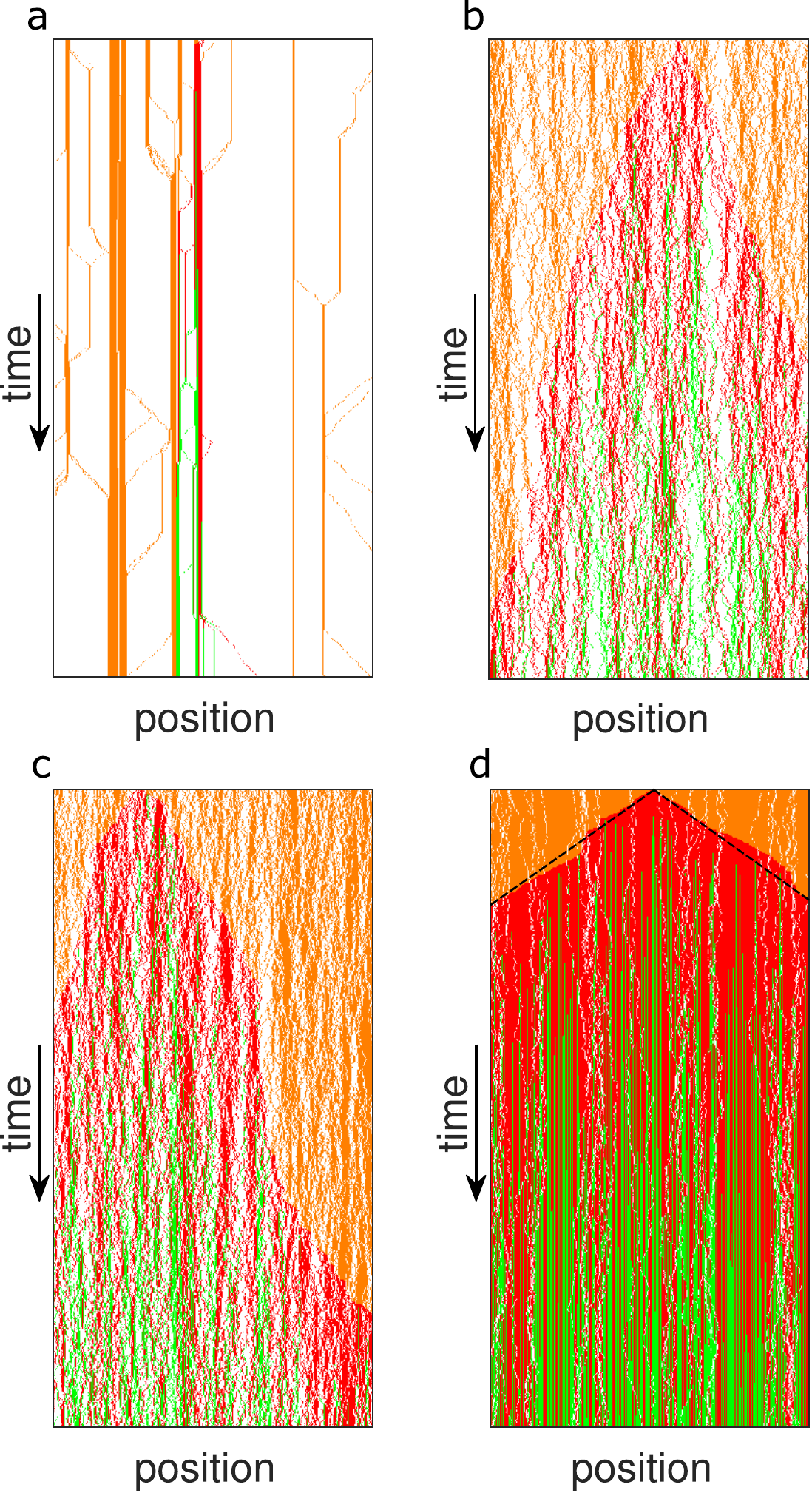}%
	\caption{Space-time evolution of clustering and contagion for $r_{\rm rec}=10^{-3}$ and (a) $\phi=0.1$ and $\nu=0.005$, (b) $\phi=0.3$ and $\nu=0.5$, (c) $\phi=0.5$ and $\nu=0.5$, and (d) $\phi=0.9$ and $\nu=0.5$. The color scheme is as follows: white corresponds to an empty site, orange is susceptible, red is infected, and green is recovered. In each panel, there are $1000$ time steps (after the beginning of the epidemic) shown along the vertical axis and $500$ lattice sites along the horizontal axis. In (d), dashed straight lines help visualize the space-time cone of infection.}
	\label{fig:cone}
\end{figure} 

The CSD, defined as the average number of clusters of size $l$ (measured in number of particles), is denoted by $F_\text{c}(l)$ and arises when the steady state is reached (within which particles continue to escape and join clusters in both right and left directions in a dynamical equilibrium, while cluster sizes fluctuate slightly). The distributions of cluster and `gas' region (i.e., empty region) sizes were shown to be well described by \cite{soto2014run}
\begin{equation}
F_\text{c}(l)=A_\text{c}\exp(-l/L_\text{c}), \,\,\,\, F_\text{g}(l)=A_\text{g} \exp(-l/L_\text{g}),
\label{eq:Fc}
\end{equation}
where the average cluster and gas region sizes are
\begin{equation}
	L_\text{c} = \sqrt{\frac{2v\phi}{\nu(1-\phi)}},\,\,\,\,L_\text{g} = \sqrt{\frac{2v(1-\phi)}{\nu\phi}},
	\label{eq:lclg}
\end{equation}
and the prefactors are 
\begin{equation}
A_\text{c} = \frac{N\nu\left(1-\phi\right)}{2v},\,\,\,\,A_\text{g} = \frac{N\nu\phi}{2v}.
\label{eq:Ac}
\end{equation}
Expressions \eqref{eq:Fc}--\eqref{eq:Ac} show excellent agreement with simulations, provided that $\phi\gg\nu/v$ but not too high. For parameters leading to the presence of a large number of very small ``clusters'' such as isolated particles, expressions \eqref{eq:Fc}--\eqref{eq:Ac} do not work so well. The consequences of such less good agreement are discussed and strongly mitigated below. For most parameter sets, there are no major consequences.

\section{SIR dynamics}
\label{sec:SIRdynamics}

We now introduce and simulate the SIR dynamics. At time step $10^7$, a particle is randomly selected to become infected, that is, after stationary clustering has been reached. Then, within the same time step, a new particle selection occurs. If the selected particle is infected, its susceptible neighbors become infected. With probability $r_\text{rec}$, the selected particle recovers. If it is susceptible and any of its neighbors is infected (this is obviously never the case when the total number of infected particles is one), the selected particle becomes infected. As before, a new director may be drawn. After that, if the director points to an empty site, the particle moves. Random particle selections continue until $M$ particles are selected within the same time step. The whole procedure is repeated for the subsequent time steps until the epidemic ends, i.e., the number of infected individuals becomes zero. The SIR variables counting the total number of particles in each contagion state are denoted by $S$ (susceptible), $I$ (infected), and $R$ (recovered). They obey the relation ${S+I+R=M}$ for the total number of particles. The population fractions are $f_{\rm X}\equiv X/M$, where $X=S,I,R$. Therefore, $f_{\rm S}+f_{\rm I}+f_{\rm R}=1$. Even though transmission is deterministic and guaranteed when the selected particle and a neighbor are in the susceptible and infected states, the effective transmission is stochastic because particle selection is stochastic. A high transmissibility regime therefore arises because typically most particles will be selected within a time step.

Fig.~\ref{fig:cone} shows the temporal evolution of the spatial configurations, affected by the clustering-epidemic coupling. The epidemic spread generates a space-time ``cone'' of infection (see Fig.~\ref{fig:cone}d), roughly speaking, as the disease starts to spread to adjacent clusters. The smaller the average gas region size, the easier to perceive the cone. An epidemic which is spreading fast across the agglomerates corresponds to a small average cone slope. The slope is determined by the tumbling plus travel time between clusters and the time needed for the infection to spread within a cluster. We numerically found that the latter increases linearly with cluster size. As time progresses, recovered particles start to appear within the cone.

\begin{figure}[!h]
	\centering
	\includegraphics[width=\columnwidth]{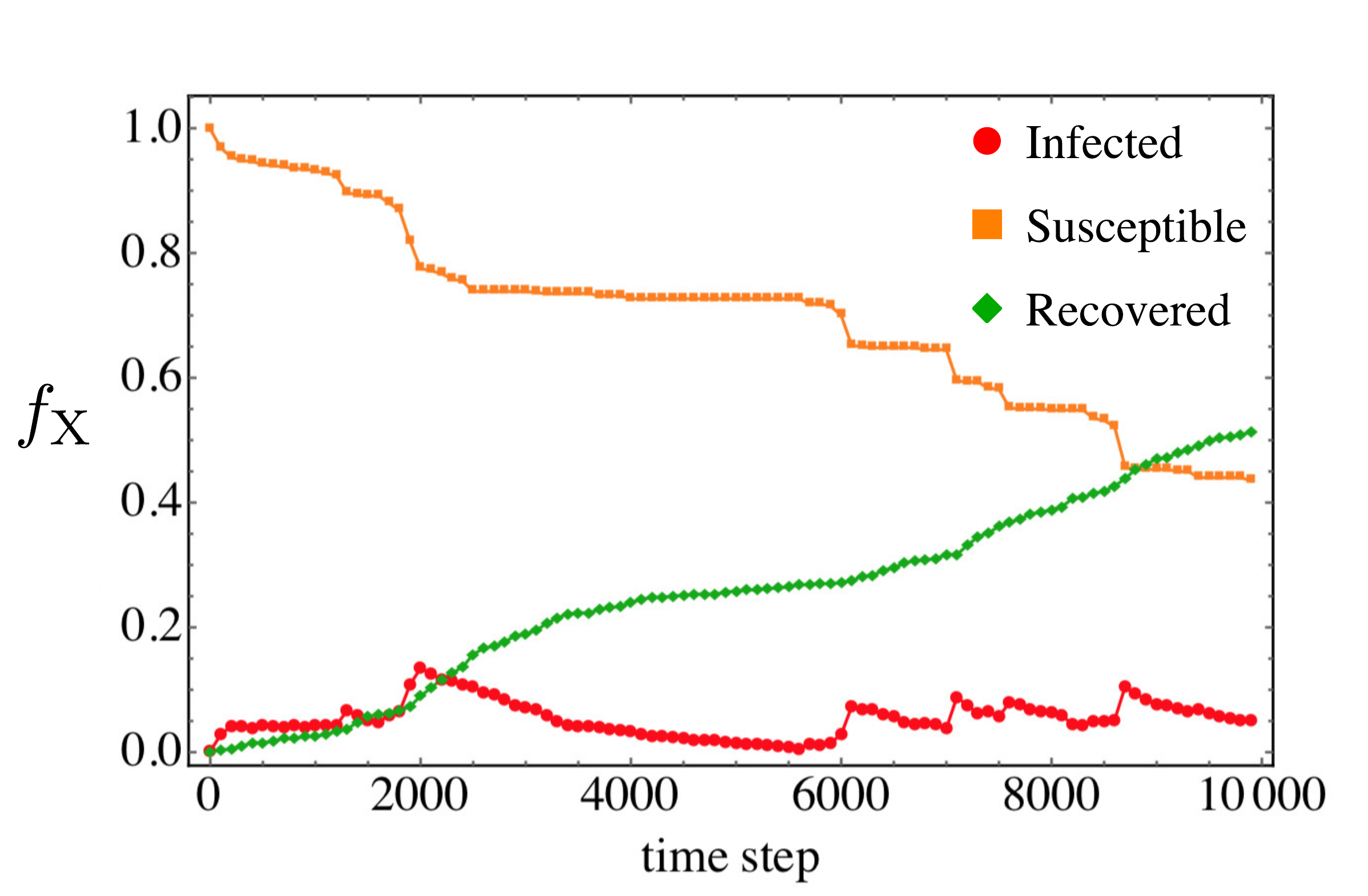}%
	\caption{ Single-realization population fraction of individuals, $f_{\rm X}$, of each type $X=S,I,R$, that is, susceptible (orange square points), infected (red circle points), and recovered (green diamond points), for each time step after the beginning of the epidemic until its end ($I=0$). Parameters: $\phi=0.4$, $\nu=10^{-3}$, and $r_{\rm rec}=10^{-3}$.}
	\label{fig:singlerealizationpopfrac}
\end{figure}

 The evolution of the population fractions is shown in Fig.~\ref{fig:singlerealizationpopfrac}. The susceptible fraction $f_{\rm S}$ decreases in a sequence of jumps resulting from infected particles arriving in susceptible clusters and infecting its individuals. The infected fraction $f_{\rm I}$ shows a corresponding series of peaks, with $f_{\rm I}$ relaxing after each peak as particles are still travelling between clusters, until a new susceptible cluster is hit. This relaxation is a result of the exponential temporal decay arising from recovered individuals. (Actually, this picture gets averaged over right and left sides.) The recovered fraction $f_{\rm R}$ always increases as particles cannot become re-infected.

\begin{figure}[!h]
	\centering
	\includegraphics[width=\columnwidth]{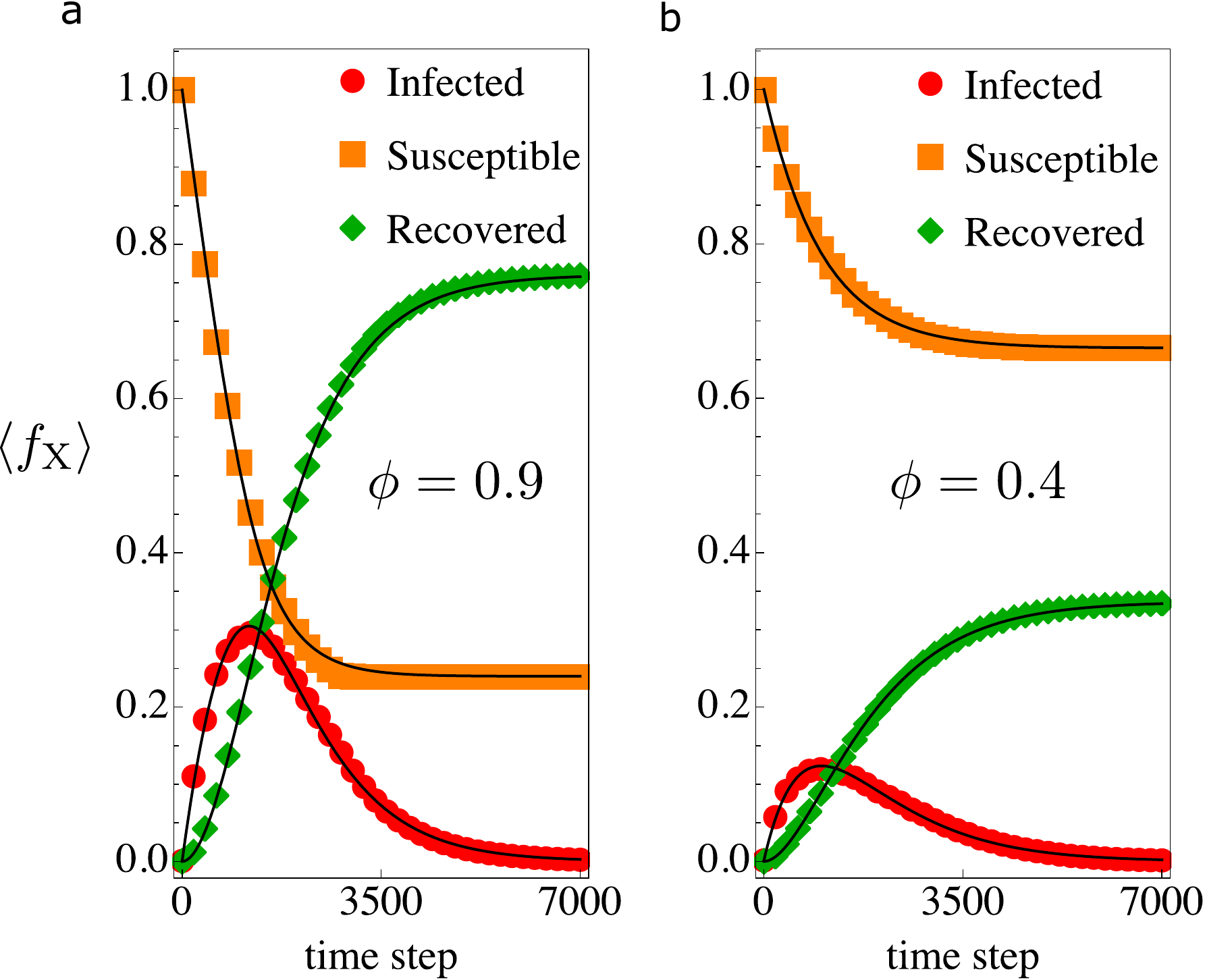}%
	\caption{Realization-average population fraction of individuals, $\langle f_{\rm X}\rangle$ (later in the text also denoted just by $f_{\rm X}$), of each type $X=S,I,R$, that is, susceptible (orange square points), infected (red circle points), and recovered (green diamond points), for each time step after the beginning of the epidemic until its end. (a) $\phi=0.9$ and (b) $\phi=0.4$. Other parameters: $\nu=0.1$ and $r_{\rm rec}=10^{-3}$. Solid black lines correspond to fitting from the theory in Eq.~\eqref{phenontheory}, with fitting parameters (a) $a=1.583\times 10^{-3}$, $\beta=1.0284$, and $S_{\infty}=432$ and (b) $a=1.493\times 10^{-3}$, $\beta=0.6803$, and $S_{\infty}=532$.}
	\label{fig:avgrealizationpopfrac}

\end{figure}

The realization averages $\langle f_{\rm X}\rangle$ are shown in Fig.~\ref{fig:avgrealizationpopfrac}. They are obtained by averaging $f_{\rm X}$ over $10^3$ numerical experiments, each starting from random homogeneous position configurations. That is, in each experiment the clusters end up forming in different positions, meaning that the travels and thus the epidemic peaks will mostly occur in different time instants. As a result, the dynamics gets smoothed out, and only a single averaged peak arises, corresponding to the average instance of cluster-cluster travels during which the particle becomes recovered before reaching a new cluster. From now on, we drop the brackets in the notation for the realization averages, i.e.,  their notation are replaced according to $\langle X\rangle\rightarrow X$ and $\langle f_{\rm X}\rangle\rightarrow f_{\rm X}$.

 We now develop a phenomenological theory for the average epidemic data in Fig.~\ref{fig:avgrealizationpopfrac}. It can be framed in terms of either $X$ or $f_{\rm X}$. First, at early times far from the (average) end of the epidemic, the behavior is well captured by the simple differential equations
\begin{equation}
	\frac{dS}{dt}=-\beta, \,\,\frac{dI}{dt}=\beta-r_\text{rec}I, \,\,\frac{dR}{dt}=r_\text{rec}I
	\label{eq:phenearly}
\end{equation}
where time $t$ is measured in time step units (but taken as continuous in the theory) and $\beta$ and $r_{\rm rec}$ are positive constants. We highlight the fact that $dS/dt$ is negative and has constant magnitude $\beta$. This constant is similar to the infection rate constant $\beta_{0}$ in traditional SIR models (which implicitly assume a 2D scenario) where $dS/dt = - \beta_{0} SI$ \cite{kermack1927contribution}. However, in our case, $\beta$ does not multiply $S I$ since each susceptible particle can be in direct contact with at most one infected, even when there is a large fraction of infected individuals throughout the system. That is, there is a ``thermodynamically'' vanishing number of possible contacts between infected and susceptible individuals. As a result, our $\beta$ has different dimensions than $\beta_{0}$. The reason for the constant behavior of $dS/dt$ stems from the truly 1D nature of the problem. The size of the infection-front is constant, i.e., the increase of $I$ does not lead to an increase in the effective infection rate since all infected particles but two (the right and left infection-front leading ones) cannot be in contact with susceptible particles, by definition. In other words, here the effective infection rate does not depend on encounters between $S$ and $I$ individuals and therefore does not depend on $S I$ (hence why the traditional basic reproduction number ``$R_0$'' of SIR models is not studied here \cite{van2017reproduction}). Also, even though the infection may be detained in some realizations because the infection-front leading particles may have not yet reached a new cluster, in other ones it will have done so, and therefore the infection will be advancing, allowing for a constant (instead of fluctuating) $dS/dt$ on average. The first term on the right-hand side of the equation for $dI/dt$ gives the corresponding increase in $I$ whereas the second term gives a simple exponential decay of infected individuals into recovered ones. Finally, the opposite of the latter term appears in the equation for $dR/dt$, concluding the set of equations for the initial regime. 

As initial conditions, we can use $S({t=0})=M$, $I({t=0})=0$, and $R({t=0})=0$. Notice that we do not need to set $I({t=0})>0$ in order to ``get things rolling'' as would be the case in a traditional SIR model. This is because our average effective rate of infection is modelled as constant (${dS/dt=-\beta}$) and does not depend on $I$, being non-zero for ${I=0}$. This is a very good approximation as long as the typical total population is much larger than one individual, which has to be the case anyway for this kind of statistical approach to work. Of course, in the agent-based simulations we need to have at least one infected. The solution of Eqs.~\eqref{eq:phenearly} is $S(t)=M-\beta t$, $I(t)=\beta/r_{\rm rec} + (1-\beta/r_{\rm rec})\exp(-r_{\rm rec} t)$, and $R(t)=M-S(t)-I(t)$ as can be readily verified. This solution works only for the early-time dynamics and, with it, the average epidemic never ends; instead, $I(t)$ would reach a plateau.

Once $S$ approaches its final value $S_{\infty}$, where we define $X_\infty\equiv X(t\rightarrow\infty)$ for $X=S,I,R$, the derivative $dS/dt$ has to go to zero. The mechanism behind it is that more and more epidemic realizations start to reach their end since infection-front leading particles eventually recover before reaching a new susceptible cluster, typically far from the extreme scenario where the whole population gets infected. Therefore, the average behavior starts to smoothly crossover from the regime with negative constant $dS/dt$ towards the regime with zero $dS/dt$. Such final smooth saturation process can be modelled using a hyperbolic tangent as follows:
\begin{equation}
\frac{dS}{dt}=-\beta \tanh{\left[a(S-S_{\infty})\right]},
\label{eq:phenS}
\end{equation}
\begin{equation}
\frac{dI}{dt}=\beta \tanh{\left[a(S-S_{\infty})\right]} -r_\text{rec}I
\label{eq:phenI}
\end{equation}
\begin{equation}
\frac{dR}{dt}=r_\text{rec}I
\label{eq:phenR}
\end{equation}
where $a$ and $S_{\infty}$ are positive constants. In Eq.~\eqref{eq:phenS}, if $S-S_{\infty}$ is high, i.e., far from the final saturation, then $\tanh{\left[a(S-S_{\infty})\right]}\approx1$ and we recover the initial regime. Another way to retrieve the initial regime is to take the limit $a\rightarrow\infty$. Therefore, the constant $1/a$ can be interpreted as a measure of the width of the crossover between the initial and final regimes, meaning that high (low) $a$ corresponds to a sharp (smooth) transition between regimes.

Other functions commonly used to model saturation processes could be used instead of $\tanh{(a x)}$, such as the apparently simpler $x/(k+x)$, with $k$ some constant. However, with the hyperbolic tangent, the analytical solution of Eqs.~\eqref{eq:phenS}-\eqref{eq:phenR} can be obtained using a symbolic calculation software and reads
\begin{equation*}
S(t)=S_\infty+\csch^{-1}\left(e^{a \beta  t} \csch\left[a (M-S_\infty)\right]\right)/a,
\end{equation*}
\resizebox{1\columnwidth}{!}{
	\begin{minipage}{\columnwidth}
		\begin{align*}
		I(t)=\frac{\beta}{r_{\rm rec}}  \left[\, _2f_1\left(\frac{r_{\rm rec}}{2 a \beta };-e^{2 a \beta t }
		\csch^2\left[a (M-S_{\infty})\right]\right)\right.\\\\\left.-e^{-r_{\rm rec} t} \, _2f_1\left(\frac{r_{\rm rec}}{2 a \beta };-\csch^2\left[a (M-S_{\infty})\right]\right)\right],
		\end{align*}
\end{minipage}
}\\
\begin{equation}
R(t)=M-S(t)-I(t),
\label{phenontheory}
\end{equation}
where $_2f_1\left(\frac{r_{\rm rec}}{2 a \beta };x\right)\equiv\,_2F_1\left(\frac{1}{2},\frac{r_{\rm rec}}{2 a \beta };\frac{r_{\rm rec}}{2 a \beta }+1;x\right)$ and
$_2F_1$ is the Gauss hypergeometric function \cite{arfken1999mathematical}. In the limit $a\rightarrow\infty$, the solution of the early-time regime is also retrieved. 
Such phenomenological theory gives very good results; see Fig.~\ref{fig:avgrealizationpopfrac}. The constants $\beta$, $a$, and $S_{\infty}$ are treated here as fitting parameters. $S_{\infty}$ can also be derived from the microscopic theory described in Section \ref{sec:micro}. Besides the data shown in Fig.\ \ref{fig:avgrealizationpopfrac}, we have successfully used Eq.~\eqref{phenontheory} to fit other data for several distinct microscopic parameter sets (data now shown), effectively exploring the whole space of parameters.

\section{Microscopic Theory}
\label{sec:micro}
We now derive a microscopic theory for the (realization average) total amount of ever-infected (recovered) particles at the end, $R_{\infty}$, where we remind that $X_\infty\equiv X(t\rightarrow\infty)$ for $X=S,I,R$. In other words, we sum up all infections over time. This will allow us to understand how the microscopic parameters, namely $\nu$, $\phi$, and $r_{\rm rec}$, affect the average aftermath impact of the epidemic. By definition, $I_\infty=0$ and thus $S_\infty+R_\infty=M$. We calculate $R_\infty$ within the approximation where all clusters have the average size $L_\text{c}$ and all gas regions have the average size $L_\text{g}$. 

Since the epidemic starts within a cluster (as even isolated particles are clusters of size one), we assume that at least one cluster becomes infected entirely, that is, the average final number of ever-infected particles $R_\infty$ is at least the average cluster size $L_\text{c}$. In this derivation, we will assume that every infected cluster will have all of its particles infected, which is true in most cases because of the high-transmissibility regime; in 1D, this is necessary for the progress of the infection, as taken into account below. Once the infection has reached a particle located at the border of the first infected cluster, the particle has to flip and travel to the neighbouring cluster in order for the epidemic to continue. Let us denote by $\tau_\text{b}$ the flip plus travel time. If during $\tau_\text{b}$ the particle recovers, the infection does not proceed. The infection will therefore proceed with probability $(1-r_\text{rec})^{\tau_\text{b}}$ on each of the left and right sides. Assuming it has reached those two new clusters, another $2L_\text{c}$ particles will be infected. The infection may then keep going, but since the process is sequential, it will do so only if the two first right and left clusters (with respect to the ``central'' initial cluster) had become infected. Thus, the probability that a third and a fourth cluster become infected (one on each side) is $(1-r_\text{rec})^{\tau_\text{b}}\times(1-r_\text{rec})^{\tau_\text{b}}= (1-r_\text{rec})^{2\tau_\text{b}}$. In summary, a geometric series arises, and $R_\infty$ can be written as
\begin{equation*}
R_\infty
=L_\text{c}+2L_\text{c}(1-r_\text{rec})^{\tau_\text{b}}+2L_\text{c}(1-r_\text{rec})^{2\tau_\text{b}}+\cdots
\end{equation*}
\begin{equation}
\hspace{-1cm}=L_\text{c}\left[1+2\sum_{k=1}^{(N_\text{c}-1)/2}(1-r_\text{rec})^{k \tau_\text{b}}\right],
\label{eq:rinf}
\end{equation}
where, on the right-hand-side of Eq.~\eqref{eq:rinf}, there are $(N_{\rm c}-1)/2$ terms after the first one, $N_{\rm c}$ being the global amount of clusters; we subtract one because the first cluster has already been counted and then we divide by two because the next clusters are counted two by two (left and right).

Using the geometric series formula, the resulting expression for $f^\infty_\text{R}=R_\infty/M=R_\infty/N \phi$, the population \textit{fraction} of ever-infected individuals, is
\begin{equation}
f^\infty_\text{R}=\frac{L_\text{c} \left[\left(1-r_{\text{rec}}\right){}^{\tau
		_b}+1-2 \left(1-r_{\text{rec}}\right){}^{\frac{1}{2} (N_\text{c}+1) \tau _\text{b}}\right]}{  \left[1-\left(1-r_{\text{rec}}\right){}^{\tau _\text{b}}\right]N \phi}
	\label{eq:rinf2}
\end{equation}
where $L_\text{c}$ and $N_\text{c}$ are known functions of the microscopic parameters $v$, $\nu$, $\phi$, and $N$, with $L_\text{c}$ given in Eq.~\eqref{eq:lclg} and the total number of clusters $N_\text{c}$ obtained by summing the CSD $F_\text{c}(l)$ in Eq.~\eqref{eq:Fc} over $l$ from ${l=1}$ to ${l=M=N\phi}$. However, Eq.~\eqref{eq:rinf2} cannot yet be expressed only in terms of microscopic parameters because the time to tumble and travel between clusters, $\tau_\text{b}$, is yet to be derived. In a first approximation, $\tau_\text{b}$ can be calculated as $\tau_\text{b}=\tau_\text{f}+\tau_\text{t}=2/\nu+L_{\rm g}/v$, where we consider the flip time $\tau_\text{f}=2/\nu$, that is, the time needed for a change of self-propulsion direction, plus the travel time $\tau_\text{t}$, which is $\tau_\text{t}=L_\text{g}/v$ if we assume ballistic motion between clusters. 
A better approximation for $\tau_\text{t}$ is to consider the diffusive nature of the travel. To do that, we take the mean-squared displacement (MSD) of a RTP in 1D in terms of $\nu$ and $v$, which reads $\text{MSD}(t)\equiv\langle x^2(t) \rangle=(v^2/2\nu^2)\left(2\nu t- 1+e^{-2 \nu  t}\right)$, where $x$ is the particle position measured in lattice sites relative to the initial position \cite{malakar2018steady}, which in our case is the cluster border, and $\langle\cdots\rangle$ indicates averaging over multiple realizations of the diffusive trajectory. This expression can be readily inverted, expressed in terms of the non-elementary function product log, and then used to calculate the corresponding diffusive travel time by identifying $\langle x^2(t=\tau_\text{t}) \rangle=L_\text{g}^2$. Therefore, the new travel time is $\tau_\text{t}= \text{MSD}^{-1}(L_\text{g}^2)$, which replaces the ballistic travel time in the calculation of $\tau_\text{b}$ above:
\begin{equation}
\tau_\text{b}=2/\nu+\text{MSD}^{-1}(L_\text{g}^2).
\label{eq:taub}
\end{equation}

\begin{figure}[!h]
	\centering
	\includegraphics[width=\columnwidth]{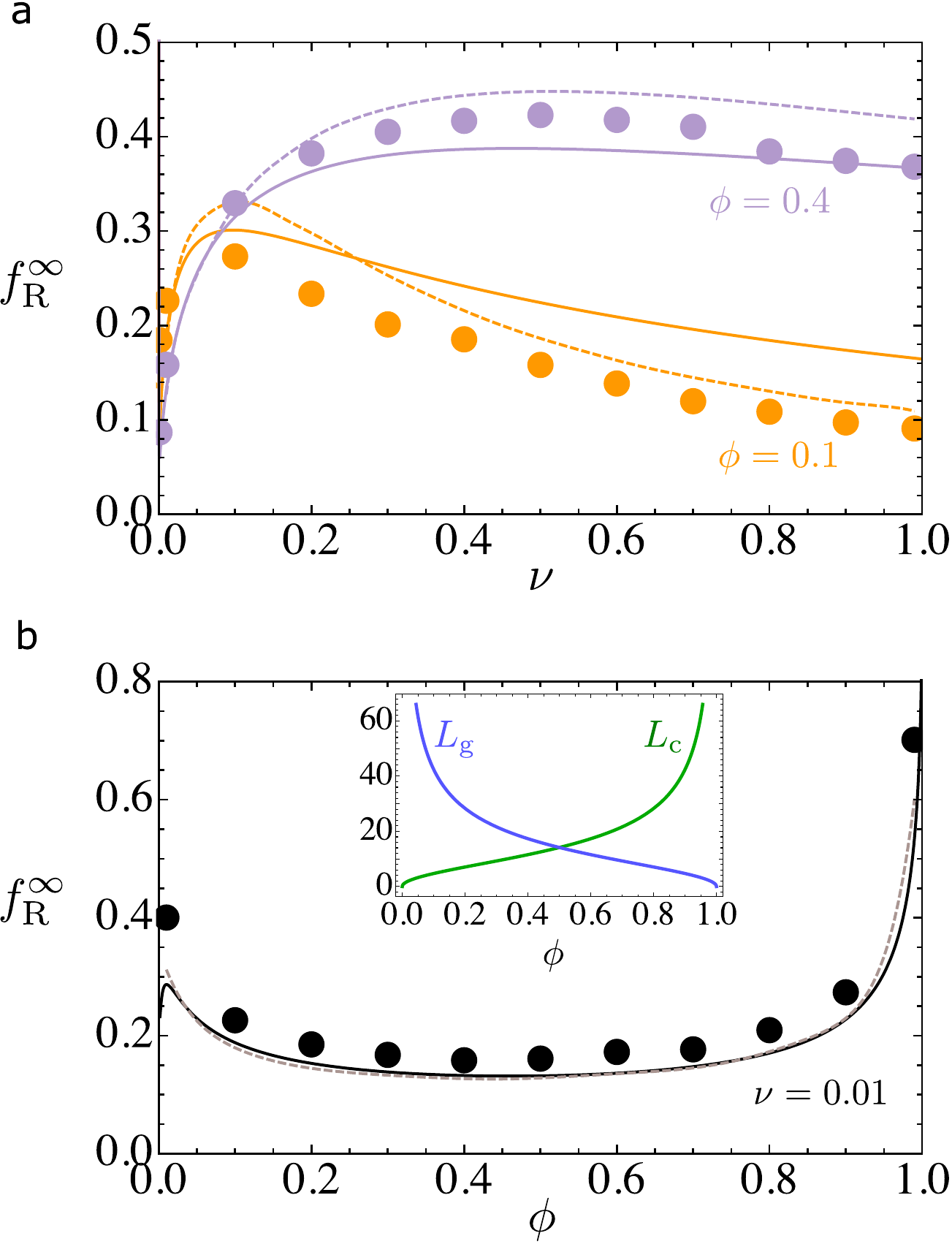}%
	\caption{Total final fraction of ever-infected individuals, $f^\infty_\text{R}$, for $r_{\rm rec}=10^{-3}$. (a) $f^\infty_\text{R}$ versus $\nu$ for $\phi=0.4$ (light purple) and $\phi=0.1$ (bright orange) and (b) $f^\infty_\text{R}$ versus $\phi$ for $\nu=10^{-2}$, with an inset showing $L_\text{c}$ (solid green line) and $L_\text{g}$ (solid light blue line) as functions of $\phi$ as given by Eq.~\eqref{eq:lclg} for $v=1$ and $\nu=10^{-3}$. The solid lines in the main plots correspond to the theoretical expression \eqref{eq:rinf2}, taking as input $L_\text{c}$, $L_\text{g}$ and $N_\text{c}$ calculated from the clustering theory in Eqs.~\eqref{eq:Fc} and~\eqref{eq:lclg} as developed in Ref.~\cite{soto2014run}. The dashed lines in the main plots correspond to the theoretical expression \eqref{eq:rinf2}, but now taking as input $L_\text{c}$, $L_\text{g}$ and $N_\text{c}$ calculated directly from the simulations, interpolated between the chosen simulation parameters.}
	\label{fig:totalfraction}
\end{figure}

Since we also know $L_\text{g}$ from Eq.~\eqref{eq:lclg}, the final expression for $f^\infty_\text{R}$ [\eqref{eq:rinf2} and~\eqref{eq:taub}] now depends only on microscopic parameters and is plotted in Fig.~\ref{fig:totalfraction} as a function of $\nu$ and then as a function of $\phi$. Note the non-monotonic dependencies, well capture by our parameter-free theory, which in turn provides a good overall agreement with simulations. An alternative approach, instead of using \eqref{eq:Fc}--\eqref{eq:Ac} for the clustering observables $L_\text{c}$, $L_\text{g}$, and $N_\text{c}$, is the following: take the \textit{measured} $L_\text{c}$, $L_\text{g}$, and $N_\text{c}$ directly from the simulation and plug them into our expression for $f^\infty_\text{R}$. By doing so, the quantitative agreement becomes even better as also shown in Fig.~\ref{fig:totalfraction}. This has to do with the fact that Eqs.~\eqref{eq:Fc}--\eqref{eq:Ac} underestimate the number of isolated particles, as previously noticed \cite{soto2014run}. When the number of such isolated particles becomes large (e.g., for large $\nu$), the clustering theory is less accurate. In summary, our contagion theory leads to satisfactorily accurate results particularly when accurate clustering observables are used as input.

\begin{figure}[!h]
	\centering
	\includegraphics[width=0.95\columnwidth]{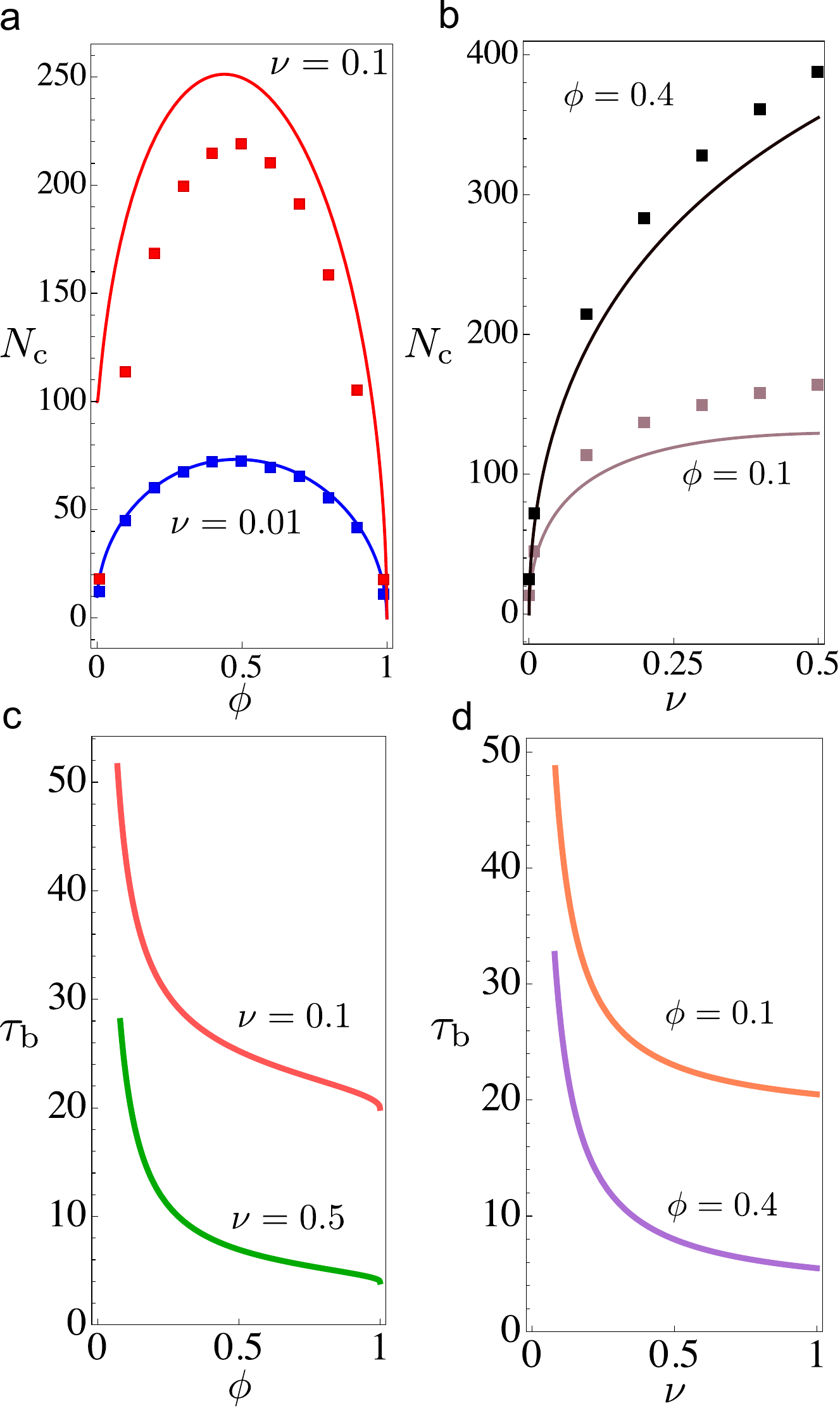}%
	\caption{(a) $N_\text{c}$ versus $\phi$ for ${\nu=0.1}$ (top) and ${\nu=0.01}$ (bottom) and (b) $N_\text{c}$ versus $\nu$ for ${\phi=0.4}$ (top) and ${\phi=0.1}$ (bottom), as measured directly from simulations (markers) and from the theory in Eqs.~\eqref{eq:Fc}--\eqref{eq:Ac} (solid lines). (c) $\tau_\text{b}$ versus $\phi$ for ${\nu=0.1}$ (top) and ${\nu=0.5}$ (bottom) and (d) $\tau_\text{b}$ versus $\nu$ for ${\phi=0.1}$ (top) and ${\phi=0.4}$ (bottom), as from the theory in Eq.~\eqref{eq:taub}. For (c) and (d), a comparison with simulation data is indirectly made in Fig.~\ref{fig:totalfraction}.}
	\label{fig:nctaub}
\end{figure}

To understand the non-monotonic behavior of $f^\infty_\text{R}$ versus $\nu$ in Fig.~\ref{fig:totalfraction}a, let us focus on the curve for $\phi=0.1$. For low $\nu$, $f^\infty_\text{R}$ is almost zero. In this limit, the clusters and the distance between them are large: particles agglomerate into a small number of large clusters due to their strong persistent motion. In this scenario, particles take a long time to flip and escape from the clusters. Therefore, particles typically recover from the infection before either leaving the cluster or reaching a new cluster. As a result, only a small fraction of individuals becomes infected in spite of the fact that they are strongly clustered. By increasing the tumbling rate $\nu$ just a little, the clusters become smaller, but the particles now take significantly less time to flip and travel [notice the $\nu^{-1}$ dependence in Eq.~\eqref{eq:taub}] and $N_\text{c}$ increases (see Fig.~\ref{fig:nctaub}b, d), both of which contribute to infections. This is enough for the particles to typically remain infected until reaching the next clusters. As a consequence, a much larger fraction of individuals gets infected. That is, the data in Fig.~\ref{fig:totalfraction}a shows that, even though the clusters are smaller, with less individuals getting infected in each cluster infection, this is not sufficient to counterbalance particles reaching new clusters sooner and the increased number of clusters. However, by further increasing the tumbling rate, the fraction of infected individuals eventually starts to decrease. This is because, besides clusters becoming really small, particles become highly diffusive, meaning that the disease cannot advance in a quasi-ballistic regime as before. In fact, $\tau_\text{b}$ versus $\nu$ approaches a finite plateau instead of decreasing to zero (a feature which cannot be captured by the ballistic approximation). At the same time, the increase in $N_\text{c}$ starts to saturate. This opens the way for the effect of a reduced $L_\text{c}$ to dominate since, for high $\nu$, $L_\text{c}$ approaches zero instead of a finite plateau. Thus, the fraction of infected individuals decreases. The same behavior holds for the curve $\phi=0.4$, except that now the peak in $f^\infty_\text{R}$ versus $\nu$ is higher and gets delayed to higher tumbling rates $\nu$.

The density dependence, $f^\infty_\text{R}(\phi)$, is also captured by the microscopic theory, as shown in Fig.~\ref{fig:totalfraction}b. One would expect $f^\infty_\text{R}$ to increase monotonically with $\phi$ since, at first glance, a higher concentration of individuals should always lead to more infections per individual. While this is exactly what occurs for higher tumbling rate $\nu$ (data not shown), for low $\nu$ as in Fig.~\ref{fig:totalfraction}b, the infected population fraction, $f^\infty_\text{R}(\phi)$, counterintuitively develops a minimum located at intermediate $\phi$.
To examine this non-monotonic behavior in $f^\infty_\text{R}$ versus $\phi$, notice that there is a $N \phi$ in the denominator since we are calculating the population fraction for a fixed total lattice size $N$. At the same time, there is $L_\text{c}$ (multiplying all terms in the numerator), which first increases sub-linearly for small $\phi$ and then super-linearly for large $\phi$; see the inset of Fig.~\ref{fig:totalfraction}b. If we ignore the rest of the $\phi$-dependent quantities in $f^\infty_\text{R}$ and consider only the behavior of $L_\text{c}/\phi$, a similar minimum appears at $\phi=0.5$.
Therefore, we are led to the conclusion that, when $\nu$ is small, the contribution of $\tau_\text{b}$ and $N_\text{c}$ to $f^\infty_\text{R}$ cannot compensate the weak increase of $L_\text{c}$ for small $\phi$ as $\phi$ is increased. This makes sense: for higher $\nu$, $\tau_\text{b}$ decreases more strongly with $\phi$ and $N_\text{c}$ increases more strongly with $\phi$ (see Fig.~\ref{fig:nctaub}a,c). Finally, we notice that, $N_\text{c}$ starts to decrease with $\phi$ for $\phi>0.5$, contributing to less infections, but the effects of a super-linearly increasing $L_\text{c}$ and a decreasing $\tau_\text{b}$ are sufficient to make $f^\infty_\text{R}$ always increase in this range of $\phi$, no matter the value of $\nu$.

\section{Conclusions}
\label{sec:conc}

In this work, we studied the contagion of spontaneously agglomerating self-propelled particles in narrowly-confined scenarios. Considering run-and-tumble particles on a one-dimensional lattice, we have developed dynamical and steady-state theories that capture the essence of the epidemic spread of a SIR disease across multiple motility-induced clusters. The steady-state theory is parameter-free and satisfactorily accurate in predicting the final population fraction of ever-infected individuals as a non-trivial function of microscopic parameters. 
 
The time evolution of each epidemic realization is characterized by a series of peaks which relax during particle travels between clusters, after which a new cluster infection may occur. The realization average version of the infected curve shows initially a constant, non-exponential increase, since in 1D the rate of new infections does not increase with the number of infected. Then, the infected curve shows a peak, and finally a relaxation due to recovery. Both the susceptible and recovered curves reach a finite plateau.

Since persistent motion is the cause of agglomeration, less persistent individual motion reduces cluster sizes and cluster-cluster distances, which in turn liberates infected individuals and increases the final ever-infected population fraction. For even more erratic individual motion, clusters become too small and, additionally, the disease no longer advances ballistically enough between the agglomerates, making the overall infected population fraction decrease. While increasing the population spatial concentration always leads to a higher \textit{absolute} number of infected individuals, the population \textit{fraction} of infected individuals may decrease with concentration for low concentration and low reorientation rate, which occurs when the number of infections depends on the combination of the other factors as a sub-linearly increasing function of the concentration.

The existence of a reorientation rate that maximizes infections may in principle be observable in real systems. When individuals have a high tendency to agglomerate, these clusters are large but also separated by large distances (for a fixed total number of individuals), leading to fewer infections. This would be the case of countries with massive cities where other regions may be almost uninhabited. When individuals have a low tendency to agglomerate, the clusters will be more numerous and separated by smaller distances, but will be also small, which may decrease infections. Further studying this aspect may prove useful in terms of mitigation strategies.

The main contribution of our analysis is to illustrate the kind of analytical approach taken here towards understanding non-trivial motion and spatial effects in epidemic-like spreading across agglomerates. We also strongly highlight that, while the contagion-clustering coupling investigated here occurs in 1D and is thus purely sequential, which must limit the infection pathways of the contagion dynamics, the behavior and mechanisms described here are already very rich and the present work constitutes an important step towards the 2D case. 

In principle, additional improvements could be made to the theoretical approaches. In the phenomenological approach to the dynamical evolution, all three constants may, in principle, be derived microscopically. The microscopic theory itself could use the entire distributions to calculate the effects of the travel time rather than only the average gas size. Those improvements, however, are not necessary to capture the essence of the clustering-epidemic coupling, as shown above. This is particularly true when our contagion theory refrains from using the clustering theory from Ref.~\cite{soto2014run} and receives as input the measured values of the clustering observables. The latter approach is a natural one: the clustering state of a group of individuals is usually accessible.

There are other mechanisms that could generate multiple stationary agglomerates instead of self-propulsion, such as in systems of non-living particles with passive (or ``equilibrium'') short-range attraction and \textit{shorter}-range repulsion (excluded volume) combined with long-range repulsion \cite{archer2008two}. In this case, notice that both of our contagion theories would still apply, where the input clustering and motion observables $L_\text{c}$, $N_\text{c}$ and $\tau_\text{b}$ would come from elsewhere as determined by the new clustering mechanism and motion rules. Therefore, our theory goes beyond active matter and is expected to apply to a wider range of multi-agglomerate spatial contagion problems.

Future directions include incorporating 2D effects as well as other epidemic processes such as exposure, re-infection, death \cite{vasconcelos2021standard} and an additional contagion-clustering coupling mechanism involving a response to infection that affects clustering \cite{stockmaier2021infectious}, e.g., where contagion affects motion and therefore clustering. Other avenues include adapting our analysis to the contexts of (i) viral phage therapy against bacterial infections \cite{rodriguez2020quantitative,meynell1961phage,marchi2020statistical}, (ii) biochemical signaling between microorganisms \cite{van2014effects,simpson2021spatial}, and (iii) social consensus \cite{crokidakis2010consequence,dornelas2018impact}. Also, applications could be sought in the context of the spread of human diseases across villages separated by long distances \cite{richards2015social,rotinsulu2021spatial}.


\section{Acknowledgements}
P.d.C.\ was supported by Scholarships No.\ 2021/10139-2 and No.\ 2022/13872-5 and ICTP-SAIFR Grant No.\ 2021/14335-0, all granted by São Paulo Research Foundation (FAPESP), Brazil. P.d.C. would like to thank Ricardo Martinez-Garcia and Daniel C.~P.~Jorge for discussions. F.G.-L.\ was supported by Fondecyt Iniciaci\'on No.\ 11220683 and ANID -- Millennium Science Initiative Program -- NCN19 170D, Chile. A.N.\ was supported by Fondecyt Iniciaci\'on No.\ 11220266.


\bibliography{Active.bib}



\end{document}